\documentstyle[12pt,iopconf]{article}
\begin{document}
\include{psfig}

\title{Numerical study of the {\it glass-glass} transition in
short-ranged attractive colloids} \author{Emanuela
Zaccarelli\footnote{E-mail: emanuela.zaccarelli@phys.uniroma1.it},
Francesco Sciortino$^{1,2}$ and Piero Tartaglia$^{1,3}$} \affil{$^1$
Dipartimento di Fisica, Universit\`{a} di Roma La Sapienza, P.le
A. Moro 5, I-00185 Rome, Italy\\ $^2$ INFM-CRS SOFT, Universit\`{a}  
di Roma La Sapienza, P.zza A. Moro 2, I-00185, Roma, Italy\\
$^3$ INFM-CRS SMC, Universit\`{a}
di Roma La Sapienza, P.zza A. Moro 2, I-00185, Roma, Italy }
\beginabstract 
We report extensive numerical simulations in the {\it
glass} region for a simple model of short-ranged attractive colloids,
the square well model. We investigate the behavior of the density
autocorrelation function and of the static structure factor in the
region of temperatures and packing fractions where a glass-glass
transition is expected according to theoretical predictions.  We
strengthen our observations by studying both waiting time and history
dependence of the numerical results.  We provide evidence supporting
the possibility that activated bond-breaking processes destabilize the
attractive glass, preventing the full observation of a sharp
glass-glass kinetic transition.  
\endabstract

\section{Introduction}
The combination of theory
\cite{fabbian99,bergenholtz99,dawson00,zaccarelli01,chenpaper,merida,sperlpisa},
experiments \cite{mallamace,science02,bartsch02,malla2,mourchid} and
numerical simulations
\cite{puertas02,foffipre,zaccarelli02,sciortino03,zaccarelli04} for
short-ranged attractive colloidal systems, i.e. colloid-polymer
mixtures in which the size of the polymers is only a few percent that
of the colloidal particles, has led in the last few years to a clear
evidence of anomalous dynamical properties for dense colloidal
systems.  In particular, the existence of a reentrant region of
(super-cooled) liquid state, which is stable up to packing fractions
higher than the correspondent purely repulsive system (hard spheres),
has been established. From this liquid region one can approach
dynamical arrest in two distinct ways \cite{sciortino02}, depending on
which control parameters for the system are manipulated. Thus, an
increase of colloid density leads to the well-known hard-sphere (or
simply `repulsive') glass, while a stronger attraction, which is
experimentally realized via an increase of polymer concentration,
produces the so-called attractive glass, where colloidal particles are
stabilized by `bonding'. Moreover, an efficient competition between
these two mechanisms of arrest originates a peculiar behaviour of
dynamical correlation functions within the reentrant fluid region, in
particular a logarithmic decay for the density autocorrelation
functions and a sub-diffusive regime for the mean squared particle
displacement\cite{zaccarelli02,sciortino03,sperl}.  All these features
have been firstly predicted by the Mode Coupling Theory (MCT)
\cite{goetze91} on the basis of the existence in the control parameter
space for these systems of a higher order singularity in the solutions
of MCT equations, which entirely governs the anomalous dynamics. This
particular point in the phase diagram, named $A_4$ singularity, is
found for a particular value of the three control parameters, i.e.
temperature, volume fraction of the colloids and range of the
attractive interaction, where the liquid, the attractive glass and the
hard-sphere glass solutions all merge into a single point.  The key
parameter for the existence of such a point is the attractive range of
the potential $\Delta$ (experimentally set by the ratio of polymer
gyration radius to the colloidal radius): for $\Delta > \Delta^*$
there is no distinction between the two glassy states, while for
$\Delta < \Delta^*$ there exists a discontinuous glass-glass
transition between the two \cite{dawson00,zaccarelli01}. The
glass-glass transition terminates at an end-point, named $A_3$ beyond
which the two glasses are indistinguishable. Such kinetic glass-glass
transition is one of the most difficult MCT prediction to detect, due
to the fact that it lies entirely in a non-equilibrium region where
both experiments and simulations are highly non-trivial.

On the other hand, while there is convincing evidence, also for
molecular glass-formers, of the accuracy of MCT predictions in the
liquid regime\cite{goetzepisa}, the same does not apply for regions
where the theory predicts a {\it glass}. Indeed, the MCT glass is
commonly termed `ideal', because it arises without the inclusion of
the so-called hopping processes, i.e. further relaxing processes which
eventually allow the system to still restore ergodicity, beyond the
ideal MCT transition \cite{hopping}. This may be better explained in
the free energy landscape picture, where a glassy state is represented
as a local (metastable) minima and a hopping process corresponds to a
jump between different minima, via a low order saddle-point\cite{pel}.

In this paper, we report an extensive numerical study of the
glass-glass transition for a specific simple model
\cite{zaccarelli02}, for which the location of the MCT transition
lines and of the higher order singularity have been extensively
studied and estimated via a mapping procedure
\cite{sciortino03,sperl}.  Thereby, after explaining in detail how
this mapping is performed, we report results in the glassy region,
including static and dynamical properties of the glassy states, as
well as we compare results obtained following different histories for
the system, due to their non-equilibrium nature.  This will confirm
the importance of hopping processes in some mechanisms of
glassification, and provide more insight on the MCT applicability to
describe different mechanism of arrest. A short account of the main
results was already anticipated in a recent Letter \cite{zacca03}.

\section{Methods}
We perform event-driven Molecular Dynamics simulations for a binary
mixture of particles interacting via a narrow square-well (SW)
potential. The parameters of the mixture are chosen in order to avoid
crystallization in the system. Thus, we consider a 50:50 mixture of
700 particles of mass $m$ with diameters $\sigma_{AA}=1.2$ and
$\sigma_{BB}=1$; the hard-core diameter for the $AB$ interaction is
$\sigma_{AB}=(\sigma_{AA}+\sigma_{BB})/2$.  We fix the width of the
attractive well as $\Delta_{ij}/(\sigma){ij}+\Delta_{ij})=3\%$.
Temperature is measured in units of the well-depth $u_0$, while time
units are $\sigma_{BB}(m/u_0)^{1/2}$. 

This system was extensively studied previously by means of theory
(MCT) and simulations. The combination of both tools allowed us to
determine, within some approximations, the location of the MCT lines,
and consequently of MCT higher order singularities, for this
particular system. In this way, we obtain the best possible diagram
for studying the dynamics of the system near the $A_4$ singularity
\cite{sciortino03} and crossing the glass-glass transition
\cite{zacca03}.

\begin{figure}[ht]
        \centerline{\psfig{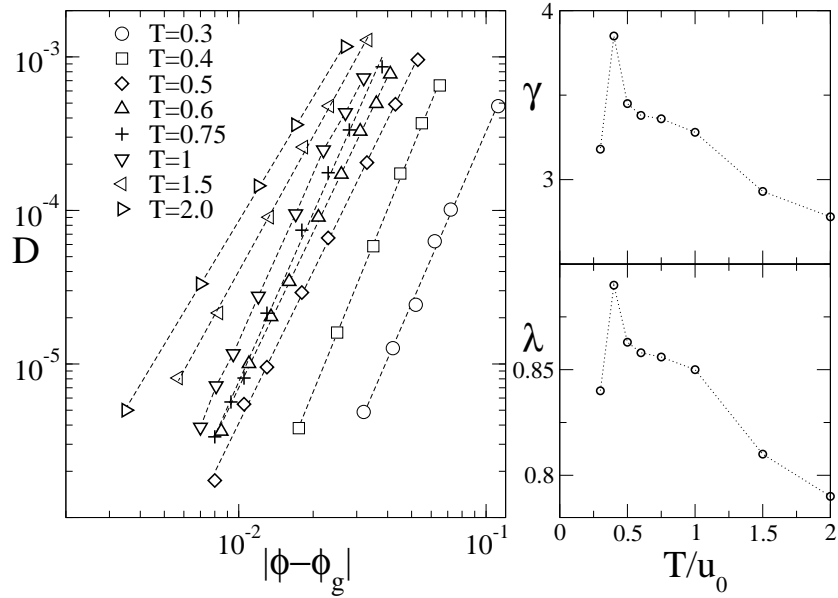} 
           } 
\caption{Left panel: Diffusion coefficients in the liquid region for
variuos studied temperatures. Dashed lines are power-law fits, with
parameters $\phi_g$, i.e. the glass transition packing fraction at
which $D=0$, and the exponent $\gamma$. Right panel: behaviour of
$\gamma$, and of the exponent parameter $\lambda$ with temperature.}
\label{fig:fits}  
\end{figure}

\begin{figure}[ht]
        \centerline{\psfig{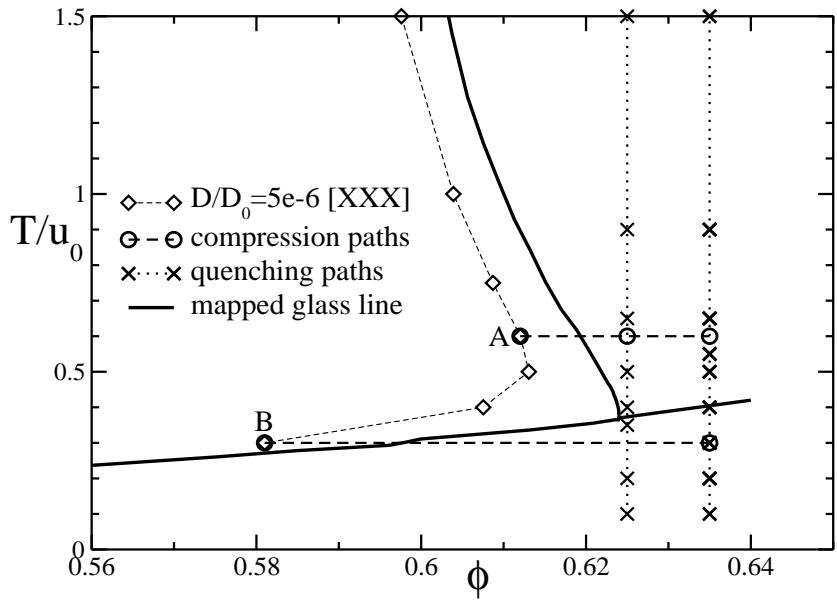} 
           } 
\caption{Map of the simulations. We consider two initial state points,
labeled $A$ and $B$, lying along the iso-diffusivity curve (diamond -
small dashed line) calculated in \protect\cite{zaccarelli02},
respectively close to the hard-sphere and to the attractive glass
transition. We compress these configurations up to two isochores,
$\phi=0.625$ and $\phi=0.635$, and follow quenches at various
temperatures (crosses - dotted lines). The solid curve is the mapped
glass line from Ref. \protect\cite{sciortino03}, as described in the
text.  }
\label{fig:map}  
\end{figure}

Here, we briefly review how this mapping procedure between theory and
simulations is performed. Firstly, we solved MCT equations for our
particular binary mixture using as inputs the partial static structure
factors calculated within Percus-Yevick (PY) approximation
\cite{konstanz}.  In particular, for the static structure
calculations, we solved numerically the Ornstein-Zernike equation on a
grid of 20000 wavevectors, with mesh 0.3141593, while for the binary
MCT equations we used a grid of 2000 wavevectors with the same mesh.
By solving the equations for a large range of $T-\phi$ values, the
PY-MCT glass lines have been detemined.  Next, we calculated the
diffusion coefficients from simulations in the liquid region, covering
4 decades in diffusivity values. A fit of these values according to
the power-law $D\sim (\phi-\phi_g(T))^\gamma$, provides estimates for
the glass transition packing fraction $\phi_g(T)$ (at which the
diffusion coefficient would vanish) for the various isotherms and a
characteristic exponent $^\gamma$ which, according to MCT predictions,
should grow near a higher order singularity. Note that $\gamma$ is
related to the exponent parameter $\lambda$ which becomes exactly
equal to $1$ at the MCT singularities ($A_3,A_4$).  Figure
\ref{fig:fits} shows the fit of the diffusivity data, together with
the parameters $\gamma$ and $\lambda$ obtained from them. Errors in
these values are given by the errors in the fits \cite{notagamma}.
The locus $\phi_g(T)$ provides the best numerical estimate of the
glass transition lines.  Next we perform a bilinear transformation in
the control parameter space $(\phi, T)$, to superimpose the PY-MCT
results for the ideal glass line with the extrapolated results from
the simulations, following an idea proposed by Sperl \cite{sperl}.
The resulting transformation is found to be,
\begin{eqnarray}
\phi_{sim} \rightarrow 1.897\ \phi_{MCT} -0.3922; \nonumber \\ 
T_{sim} \rightarrow 0.5882\ T_{MCT} - 0.225.
\label{eq:transformation}
\end{eqnarray}
This transformation provides a tool for comparing theoretical
predictions and simulation results.  Improvements over such procedure
could be obtained by using the numerical "exact" structure factors as
oppose to the PY-closure in the estimate of the ideal MCT lines.
Indeed, despite the quite good agreement of the PY solutions, small
differences between PY and numerical structure factors are crucial for
a full quantitative agreement of theory and simulations results, as
recently discussed in \cite{foffialpha}.  Assuming that the bilinear
transformation is invariant for small changes of $\Delta$, we can
locate theoretically the position of the $A_4$ singularity, moving
along the $\Delta$-axis in the control parameter space and find the
corresponding location for the real system \cite{sciortino03}. The
qualitative general agreement between predictions and simulations
results for the mean squared displacement and density auto-correlation
functions shown in Ref. \cite{sciortino03} strongly supports the
validity of the mapping procedure.  Having established the quality of
the mapping, we can locate the theoretically predicted glass-glass
transition for the $\Delta=0.03$ case which refer to the present
simulation parameters. More specifically, the glass-glass line is
found to be located in the $(\phi,T)$-plane approximately between
$(0.625,0.37)$ and $(0.64,0.41)$. The resulting liquid-glass and
glass-glass lines for the model are shown in Fig. \ref{fig:map}.

Our aim is to perform MD simulations across the glass-glass transition
line and study the dynamical properties, in particular the collective
density-density correlation functions, both in the hard-sphere and in
the attractive glass region. A discontinuity in such properties should
be observed, according to MCT, at the glass-glass transition. In order
to access the glass region, we used the following procedure, that is
schematized in Fig. \ref{fig:map}.  We consider the slowest
equilibrated liquid state point in the reentrant region,
i.e. corresponding to initial packing fraction and temperature of
$\phi_i=0.612$ and $T_i/u_0=0.6$ (labelled as $A$ in the figure). We
then rapidly compress the configuration by progressively growing the
particle radii in successive steps.  In order to investigate the
glass-glass transition we choose to compress isothermally the system
up to $\phi=0.625$ and up to $\phi=0.635$. At each of the two packing
fractions, we then suddenly change the mean square velocities to
achieve configurations at several different temperatures (indicated
with crosses).  A thermostat with a very short characteristic time
kept the temperature constant during the following aging
evolution. This procedure has been repeated for 136 independent
initial configurations extracted from the the initial equilibrium
liquid state.

Being the system out-of-equilibrium, the dynamical observables will
have a dependence also on the waiting time $t_w$, and on the chosen
initial state \cite{aging,sciotar}. Thus, in particular the density
autocorrelation functions are defined as,
\begin{equation}
\phi_{q}(t_w+t,t_w)\equiv \langle \rho^*_{\bf q}(t_w+t) \rho_{\bf
q}(t_w) \rangle/\langle|\rho_{\bf q}(t_w)|^2\rangle
\end{equation}
where $\rho_{\bf q}(t)=\frac{1}{\sqrt{N}} \sum_i e^{i{\bf q} {\bf
r}_i(t)}$ and ${\bf r}_i(t)$ are the coordinates of particle $i$ at
time $t$. The correlation functions calculated in this paper always
refer to all particles in the mixture.  Averages are taken over up to
136 independent initial configurations from the initial liquid state.
To assess the problem of a possible path-dependence of the results, we
also consider a different initial liquid state, this time within the
attractive region, i.e. $\phi_i=0.58, T_i/u_0=0.3$ (labelled as $B$ in
the figure).  Moreover, we check that our results are waiting-time
independent within a well-defined, even though narrow, time
window. This was already discussed in \cite{zacca03}.  For the present
purposes, we recall that the results reported in this article
correspond to $t^*_w=4\cdot10^3$ which provides sufficient
equilibration for the decay processes taking place on a time scale one
order of magnitude smaller \cite{sciotar}.  For this reason we only
show the decay of correlation up to $t=400$. However, in order to
provide convincing evidence that results for $t<400$ do not depend on
the chosen $t_w$, we will also report results corresponding to a $t_w$
larger by an order of magnitude.  Finally, we will discuss the
behaviour of static structure factors in the glassy region.

\section{Results}
\begin{figure}[ht]
 \centerline{\psfig{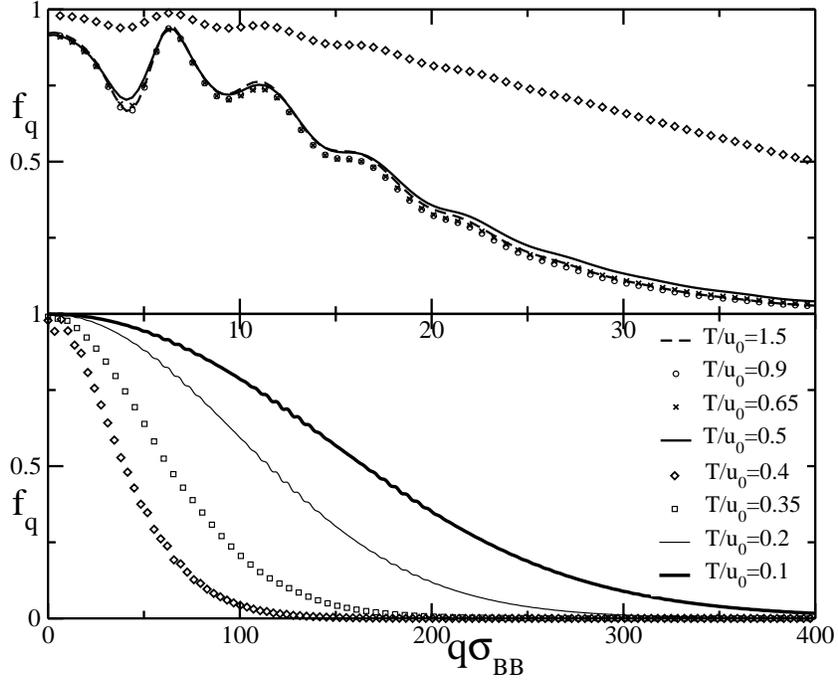} 
}
\caption{ MCT predictions for the number density non-ergodicity
parameter $f_q$ for $\phi\simeq 0.5362$ (corresponding to
$\phi_{SIM}=0.625$).  The top one refers to high temperatures, in the
hard-sphere glass region, and the bottom one to the lower ones in the
attractive glass. The case $T/u_0=0.4$ is reported in both
panels. Note the non-monotonicity of $f_q$ in the hard-sphere region,
as well as the difference of one order of magnitude in the $q\sigma$
scale betwen the two panels. }
\label{fig:fq}  
\end{figure}

Before showing the simulation results, we start by reporting MCT
predictions. In Figure \ref{fig:fq}, we show the behaviour of the
(total) non-ergodicity factor $f_q=\lim_{t\rightarrow\infty}\phi_q(t)$
(where we assume no waiting time dependence), crossing along an
isochore the glass-glass transition, at temperatures whose distance
from the transition is the same as for the simulated points.  It is
evident that $f_q$ does not change very much in the hard-sphere glass
region, but jumps abruptly at the glass-glass transition, and then
changes dramatically with temperature in the attractive region. We
here want to draw the reader's attention to the non-monotonic
behaviour of $f_q$ in the hard-sphere region with temperature, as this
non-trivial feature will be found also in the simulations.  The
crossing arises because at this isochore at intermediate temperatures
(see Fig. \ref{fig:map}), there are two distinct transitions close-by,
i.e. standard liquid-glass and glass-glass one, which can effectively
compete.  Similarly, if we plot the full time evolution of $\phi_q(t)$
at the same points for a particular wave-vector, eg. at the static
structure factor first peak, we find a net difference in the levels of
the reached plateaux, i.e. very high for the attractive glasses and
much lower for the hard-sphere ones (see Fig.2b of
ref. \cite{zacca03}).  We recall here that the glass-glass transition
is a purely kinetic phenomenon, which arises from the non-linearity of
MCT equations. Thus, there is no abrupt change in the static
structure, passing from one glass to the other.

\begin{figure}[ht] 
 \centerline{\psfig{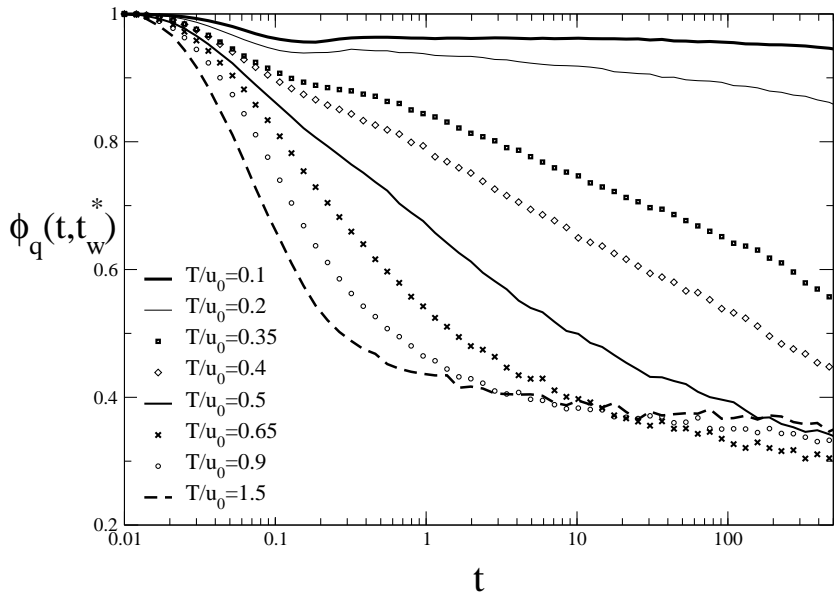}
}
\caption{ Density correlators calculated from the simulation at
$\phi=0.625$ for $q\sigma_{BB}\simeq 25$ and $t_w^*=3754$. }
\label{fig:625}  
\end{figure}

Now we turn to the simulation results.  We plot the density
autocorrelation functions $\phi_q(t,t^*_w)$ for $\phi=0.625$ at all
studied temperatures for the representative wave-vector
$q\sigma_{BB}\simeq 25$, since their behaviour is similar for all
studied wave-vectors.  The following features are evident in Figure
\ref{fig:625}:
\begin{itemize}
\item at higher temperatures, the system is already able to relax to
its plateu, even within the small studied time window. Also, the
predicted non-monotonic behaviour of the plateau with temperature,
found theoretically, is reproduced in full details;
\item lowering the temperature, there is a smooth change in the
correlators, which start to display a quite evident logarithmic decay,
signature of the higher order MCT singularity that is closeby;
\item at temperatures already below the glass-glass line, i.e. for
$T/u_0 \leq 0.4$, the correlators continue quickly to drop down, below
the expected attractive plateau; a further decrease in temperature
elongates the time duration of the plateau, but a decay at longer
times is present even in the extreme case $T/u_0=0.1$;
\item looking in more details, we observe that theoretically
$f_q(T/u_0=0.9)<f_q(T/u_0=0.65)$ (see Fig.~\ref{fig:fq}) while the
simulations show the opposite feature at our final time of
observation. This could be probably due to the inaccuracy of PY
structure factors that we used to solve MCT
equations\cite{foffialpha}, or to the fact that hopping processes
start already to produce deviations from MCT at $T/u_0=0.65$ (in this
respect, at the immediately lower temperature $T/u_0=0.5$, MCT
predictions are completely far off the simulation data).
\end{itemize}

A very similar behaviour is found analyzing the different packing
fraction $\phi=0.635$, already reported in \cite{zacca03}. The only
difference with the previous case is that no evident crossing of $f_q$
was found there. This is again in agreement with MCT, because the
influence of the liquid-glass transition is now less strong, since the
chosen isochore lies more deeply into the glassy region.

Conclusions that can be drawn from these results are that the
hard-sphere glass predictions are fully reproduced by simulations,
while the attractive glass seems to be less stable than predicted.  An
interpretation of these findings is possible taking into account a
relevant parameter for the system at low temperatures, i.e. the
lifetime of the attractive bonds. The square well model is
particularly suited to study bond dynamics, since a bond is
unambigously defined when two particles have a pair interaction energy
equal to $-u_0$, i.e. they are at a distance less or equal than
$\Delta$. To monitor the dynamical evolution of the bonds, we
introduce the `bond' correlation function $\phi_B(t_w+t,t_w)$, defined
as
\begin{equation}
\phi_B(t_w+t,t_w)=\langle \sum_{i<j} n_{ij}(t_w)n_{ij}(t_w+t)\rangle /
[N{n}_B(t_w)]
\end{equation} 
where $n_{ij}(t_w)$ is 1 if two particles are bonded and 0 otherwise,
while ${n}_B(t_w) \equiv \langle \sum_{i<j}n_{ij}(t_w) \rangle/N $ is
the average number of bonds per particle at
$t_w$. $\phi_B$ counts how many of the bonds found at
time $t_w$ are still present after time $t_w+t$,
independently from any breaking-reforming intermediate process.

\begin{figure}[ht] 
 \centerline{\psfig{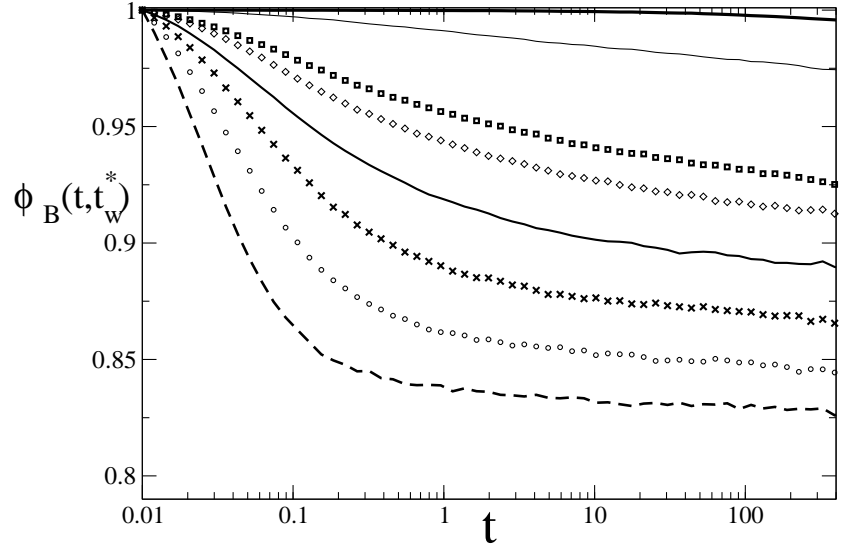}
}
\caption{ Time evolution of the bond correlators $\phi_B(t_w+t,t_w)$
for $\phi=0.635$ and $t_w^*$. Temperatures are the same as in the
previous figure (corresponding to same symbols).}
\label{fig:bonds}  
\end{figure}

\begin{figure}[ht] 
 \centerline{\psfig{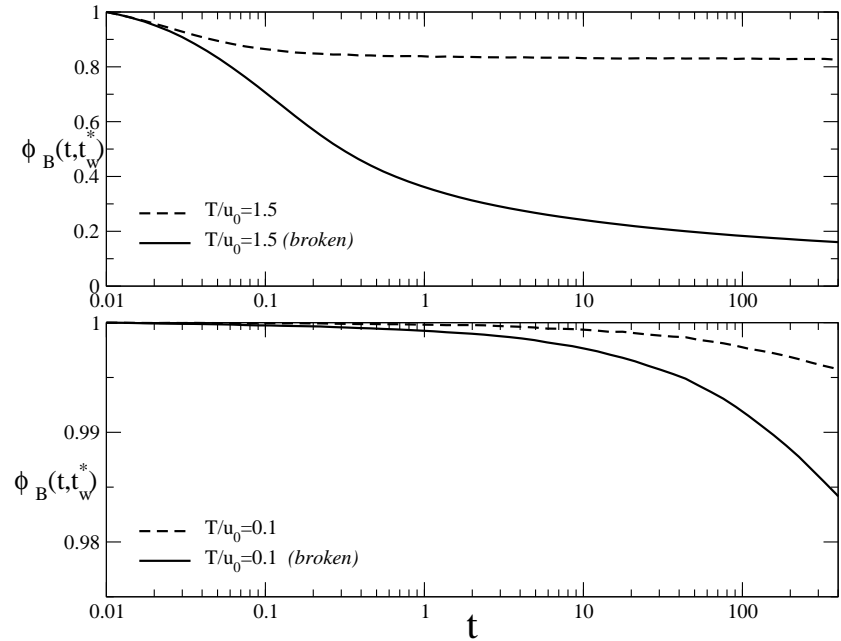}
}
\caption{ Time evolution of the bond correlators $\phi_B(t_w+t,t_w)$
for $\phi=0.635$ and $t_w^*$ for temperatures $T/u_0=1.5$ and
$T/u_0=0.1$, comparing the case when bonds after being broken are
still counted (same as \protect\ref{fig:bonds}, dashed lines) or not (curves
labelled as `broken', full lines).}
\label{fig:bonds2}  
\end{figure}

The time evolution of $\phi_B(t_w+t,t_w)$, shown in
Fig.\ref{fig:bonds}, is very similar to the one of
$\phi_q(t_w+t,t_w)$, suggesting that the decay of density-density
correlation is associated to the dynamics of bond breaking.  The
reason why the shape of the correlators in Fig.\ref{fig:bonds} looks
similar to density correlators even for high temperatures, lies in the
fact that we have taken into account also bonds re-forming, once they are broken. 
Indeed, the packing fraction we are studying ($\phi=0.635$) is well above the
glass transition packing fraction ($\phi_g \sim 0.58$) for simple hard
spheres, thus particles are confined to rattle in their cages, and
bonds gets continuously broken and reformed. To give a convincing evidence of the
different dynamics taking place at high and low temperatures, we
report in Fig. \ref{fig:bonds2} the bond correlation function,
at the two extreme temperatures $T/u_0=0.1$ and $T/u_0=1.5$, defined above, as well as 
the same functions but where $n_{ij}(t_w)$
is set to zero at the time when the $ij$-bond is broken, and kept to zero from then
on, irrespectively of the presence of the $ij$-bond at later times. 
In the high temperature case, the bond correlation function goes almost to zero
when broken bonds are not re-counted, while at low temperature there
is no substantial change.  In our previous work \cite{zacca03}, we
have already put forward the observation of relevance of bond dynamics
for attractive glasses, by showing how a characteristic type of
hopping processes takes place in the attractive glass, preempting the
observation of a sharp glass-glass transition, as predicted by MCT,
and destabilizing the attractive glass. The study of the effect of the
bonds lifetime on the full dynamical behaviour of short-ranged
attractive colloids has also been explored in \cite{saika04}.

In the present work, we support the hypothesis of bond-breaking
hopping processes preempting the observation of a sharp glass-glass
transition, as predicted by MCT, showing that results for the density
correlators as shown in Fig.~\ref{fig:625} do not depend on the chosen
waiting time or on the history path dependence.

Thereby, we first report in Fig.~\ref{fig:path} the density
correlation functions for three representative temperatures,
i.e. $T=1.5,0.35,0.1$, for a waiting time of $5\cdot 10^4$, compared
to the case examined so far for a $t_w$ smaller by a full order of
magnitude. These results are averaged over up to $40$ configurations
only, due to the very large computer time involved in the calculation
(approximately $10^3$ CPU hours for each temperature). It is evident
that no significant change in the density correlators behaviour
arises, though small discrepancies are present also due to the larger
errors present in the longer simulation runs. However, the most
important feature to underline is that an increase in waiting time
does not manifest in an increase of the time duration of the
attractive plateau. Our study shows that the stability of such
plateau, i.e. of the bond lifetime, can only be achieved by decreasing
temperature. Thus, the observed phenomenon is entirely due to bond
dynamics, which is independent on the waiting time and completely
controls the attractive glass transition.

 Of course, an interesting issue would be to know exactly where, at
infinite times, the attractive glass correlation functions would decay
to. In other words, we know that the plateau must be finite, and its
lower bound should be represented by the hard-sphere plateau. So, we
can speculate that the long-time cage of the attractive glass is still
of the order of the hard-sphere one, though at each time particles are
trapped within the attractive well-width, which is highly affected by
breaking processes, differently from simple packing.  Unfortunately,
we cannot support this picture by simulations, since it would require
many additional decades in time, beyond computational efforts.

The last point we wish to address is the dependence on the chosen
liquid initial state. Indeed, one might argue that we started our
non-equilibrium (compression+quenching) path from a state point close
to a hard-sphere glass. Thus, we can repeat the full procedure from a
different initial state close to the attractive glass transition. For
this purpose, we refer to the initial state point $B$ of Figure
\ref{fig:map}.  Results for the density correlators are compared, for
the two different initial points, in Figure \ref{fig:path}, for
temperatures $T/u_0=0.1,0.4,0.9$. It is clear that quantitatively
results are different, but qualitatively they are not. As expected,
starting from a lower temperature causes the produced glass to have a
higher correlation function, i.e. it imprints in it more of the
attractive features. However, after the microscopic time, the decay is
very similar in the two cases. Most importantly, the lowest
temperature plateau is not stabilized by the different path.

It is interesting to note that if one examines the complete
$T$-dependence of the density correlators at the same isochore
$\phi=0.625$ as in Fig.~\ref{fig:625}, but now starting from $B$, one
does not observe the same non-monotonic behaviour, at high
temperatures, for the density correlators.  This is due to the fact
that we are starting from a configuration close to the attractive
line, which somehow remains imprinted in the {\it memory} of the
system (see Figs.~\ref{fig:path},\ref{fig:sq}). Thus, probably, our
small time window of observation is not sufficient for the system to
forget about its initial state, which prevents the observation of
smaller plateaux especially at higher temperatures. In other words,
the system cannot really feel enough competition from two transitions,
and the only relevant one for the dynamics here is the glass-glass
one. Interestingly enough, this behaviour resembles closely the one at
the highest isochore ($\phi=0.635$) starting from $A$\cite{zacca03}.

Finally, we want to focus on the behaviour of the static structure
factors in the glassy region.  As explained above, $S(q)$ is the input
needed to solve MCT equations, and at the glass-glass transition it
does not have any peculiar change, other than the smooth one
characteristic of a slight change of thermodynamic control
parameters. In Figure \ref{fig:sq}, we show the (total) static
structure factors for $T/u_0=0.1,0.9$ together with the equilibrium
configuration one, both for the path starting from point $A$ (top
panel) and from point $B$ (bottom panel).  It is clear that the
structure just changes smoothly along the quenching paths, but does
change with respect to the initial equilibrium state because of the
density increase. In the glass, no detectable waiting time dependence
for the structure is observed. Static structure factors between the
two panels, corresponding to the same temperatures after different
paths, cannot be really distinguished within numerical
errors. However, we expect small changes due to the different
imprinted structures for the initial state-point. This could help
explaining why the results for the dynamical correlators are different
between the two paths (see Fig. \ref{fig:path}.).  The results shown
here for the static structure factors are quite different than those
found in a micellar system \cite{malla2}, and that could explain the
different conclusions drawn in that paper.

\section{Conclusions}
In this paper, we have reported numerical simulations along the
theoretically predicted glass-glass transition between repulsive and
attractive glass for a simple model of short-ranged attractive
colloids. A simple mapping between theory and simulations, performed
for this particular model, allowed us to have an accurate location
of this glass-glass transition, and thus, to have a consistent test of
the theoretical predictions by simulations in the glass.

Studying the glassy dynamics, we can discriminate time intervals
where waiting time does not play a significant role, and where there
is only a minor dependence on the preparation of the glassy state. It
is, therefore, an ideal system to study and compare the two mechanisms
of glassification, i.e. packing and attraction.

Our results clearly show that a glass-glass transition exists. Indeed,
there is a significant difference in the dynamical behaviour between
hard-sphere and attractive glass, though the crossover from one kind
to the other is smoother than predicted by MCT. Thus, the transition
does not manifest as sharply as predicted by the theory, due to
bond-breaking processes that reduce the stability of the attractive
glass. The detailed studies of waiting time and path dependence
strengthen this interpretation, because they do not have any effect on
the stability of the attractive plateau.  This, indeed, can only be
made longer-lasting in time by a decrease of temperature, i.e. an
(Arrhenius) increase of the bond lifetime.  Therefore, we claim that
bond-breaking activated processes have the same relevance on
attractive glass as they do in molecular glass-formers, while
hard-sphere glasses, due to their entropic origin, are basically
unaffected by them.  On the other hand, the intrinsic different nature
of the two glasses can still be experimentally detected. In
particular, for systems where the bond lifetime is long enough, or
equivalently by looking at time-scales that are shorter than the bond
lifetime, the MCT predictions for the glass-glass transition could be
fully recovered.

\section{Acknowledgments}
We acknowledge support from MIUR COFIN 2002, FIRB and INFM Iniziativa
Calcolo Parallelo. We thank S. Buldyrev for the MD code, W. G\"otze,
M. Fuchs, W. Poon and E. Bartsch for discussions.

\section*{FIGURES}



\begin{figure}[ht] 
 \centerline{\psfig{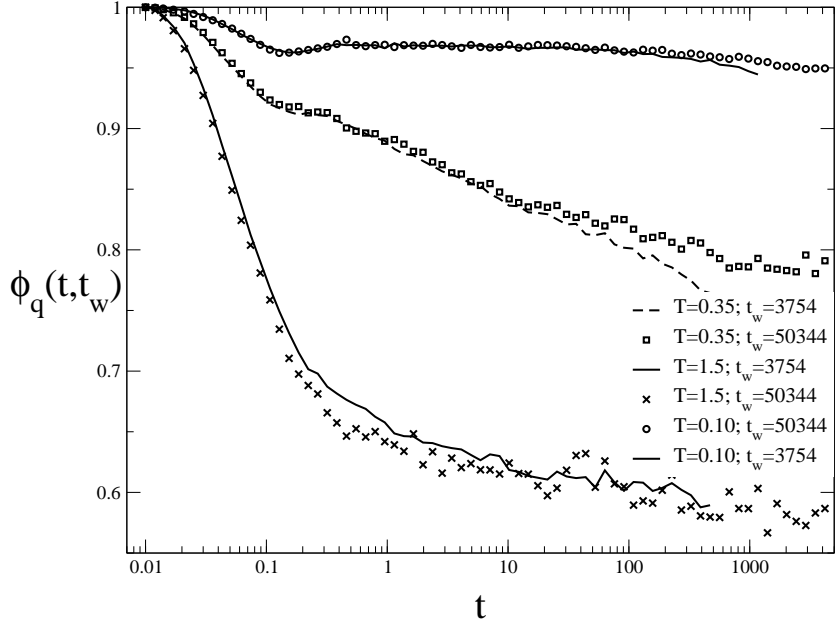}
}
\caption{ Study of the waiting time dependence for three different
temperatures at $\phi=0.635$. Points refer to $t_w=50344$, one order
of magnitude larger than $t_w^*$. Clearly, the plateaux are not
affected by the difference in waiting time, at all temperatures.}
\label{fig:errors}  
\end{figure}

\begin{figure}[ht]
 \centerline{\psfig{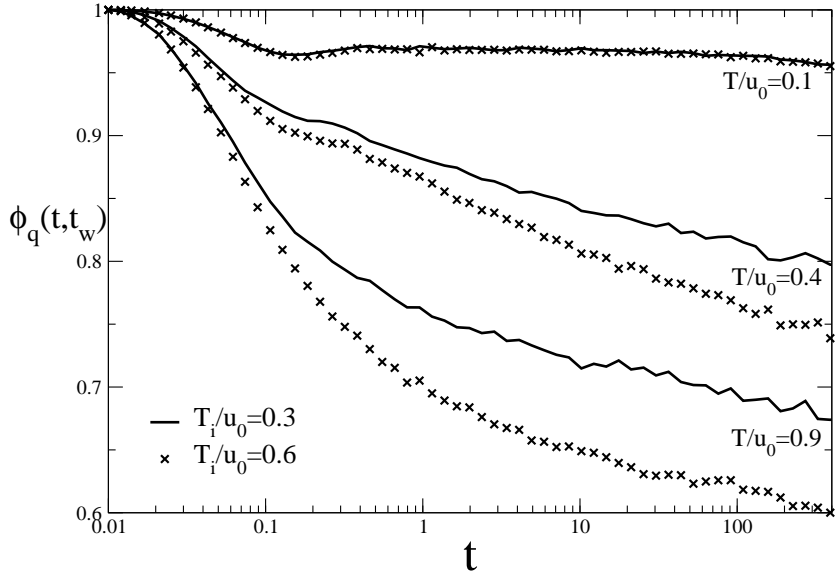}
}
\caption{ Study of the history dependence for three different
temperatures at $\phi=0.635$ and $t_w^*$. Points refer to initial
state point $B$ in Fig. \protect\ref{fig:map}, while solid lines are
for the standard initial state point $A$.}
\label{fig:path}  
\end{figure}

\begin{figure}[ht]
 \centerline{\psfig{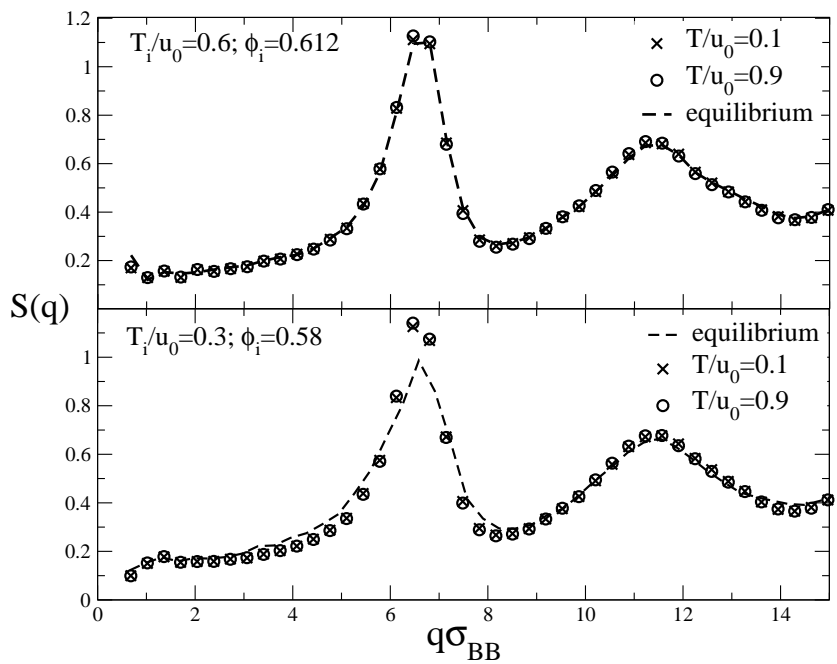}
}
\caption{ Static structure factors at different temperatures for
$\phi=0.635$, compared to each other and to the equilibrium static
structure factor of the initial liquid: top panel is for initial state
point $A$ ($T_i=0.6,\phi_i=0.612$), bottom for initial state point $B$
($T_i=0.3,\phi_i=0.58$).}
\label{fig:sq}  
\end{figure}


\begin{thebibliography}{99}

\bibitem{fabbian99} L. Fabbian, W. G\"otze, F. Sciortino, P. Tartaglia
         and F. Thiery, {\it Phys. Rev. E} R1347 (1999), and
         {\it Phys. Rev. E} {\bf 60}, 2430 (1999).

\bibitem{bergenholtz99} 
         J. Bergenholtz and M. Fuchs, {\it Phys. Rev. E }{\bf 59}, 5706 
         (1999). 

\bibitem{dawson00} K. A. Dawson, G. Foffi, M. Fuchs, W. G\"{o}tze,
        F. Sciortino, M. Sperl, P. Tartaglia, Th. Voigtmann and
        E. Zaccarelli, {\it Phys. Rev. E} {\bf 63}, 011401 (2001).

\bibitem{zaccarelli01} E. Zaccarelli, G.Foffi, P. Tartaglia,
          F.Sciortino and K. A. Dawson, {\it Phys. Rev. E} {\bf 63}, 031501
          (2001).

\bibitem{chenpaper} K. A. Dawson, G. Foffi, F. Sciortino, P. Tartaglia
  and E. Zaccarelli, {\it J. Phys.: Condens. Matter} {\bf 13}, 9113 (2001).


\bibitem{merida} K. A. Dawson, G. Foffi, G. D. McCullagh,
F. Sciortino, P. Tartaglia and E. Zaccarelli, {\it J. Phys.:
Condens. Matter} {\bf 14}, 2223 (2002).

\bibitem{sperlpisa} W. G\"{o}tze and M. Sperl,
	{\it J. Phys.: Condens. Matt.} {\bf 15}, S869 (2003).


\bibitem{mallamace} F. Mallamace P. Gambadauro, N. Micali,
        P. Tartaglia, C. Liao and S. H. Chen, {\it Phys. Rev. Lett.}
        {\bf 84}, 5431 (2000).

\bibitem{science02} 
        K. N. Pham, A. M. Puertas, J. Bergenholtz, S.U.Egelhaaf,
        A. Moussaid, P.N. Pusey, A.B. Schofield, M. E. Cates, M. Fuchs
       and W.C.K. Poon, {\it Science} {\bf 296}, 104 (2002).

\bibitem{bartsch02} T. Eckert and E. Bartsch, {\it Phys. Rev.
  Lett.} {\bf 89}, 125701 (2002); {\it Faraday Disc.} {\bf 123}, ... (2003).

\bibitem{malla2} W. R. Chen, S. H. Chen and F. Mallamace, {\it
Phys. Rev. E} {\bf 66}, 021403 (2002); S. H. Chen, W. R. Chen and
F. Mallamace, {\it Science} {\bf 300}, 619 (2003).

\bibitem{mourchid} J. Grandjean and A. Mourchid, Europhys. Lett. {\bf
65}, 712 (2004).

\bibitem{puertas02} A. M. Puertas, M. Fuchs and M. E. Cates, {\it
        Phys. Rev.  Lett.} {\bf 88}, 098301 (2002); {\it Phys. Rev. E}
        {\bf 67}, 031406 (2003).

\bibitem{foffipre}G. Foffi, K. A. Dawson, S. Buldyrev, F. Sciortino,
 E. Zaccarelli, P. Tartaglia, {\it Phys. Rev. E} {\bf 65}, 050802
 (2002).

\bibitem{zaccarelli02} E. Zaccarelli, G. Foffi, K. A. Dawson,
S. Buldyrev, F. Sciortino, P. Tartaglia, {\it Phys. Rev. E} {\bf 66},
041402 (2002).

\bibitem{sciortino03} F.  Sciortino, P.  Tartaglia, E.  Zaccarelli,
  {\it Phys. Rev. Lett.} {\bf 91}, 268301 (2003).

\bibitem{capri} E. Zaccarelli, S. Buldyrev, F. Sciortino,
P. Tartaglia, Physica A, in press (2004); cond-mat/0310765.

\bibitem{zaccarelli04} E. Zaccarelli, H. L\"owen. P.P.F. Wessels,
F. Sciortino, P. Tartaglia and C. N. Likos, condmat/0402254;
Phys. Rev. Lett., in press (2004).

\bibitem{sciortino02} 
F. Sciortino, {\it Nature Materials} {\bf 1}, 145 (2002).



\bibitem{sperl02} 
	W. G\"{o}tze and M. Sperl, {\it Phys. Rev. E} {\bf 66},
	011405 (2002).

\bibitem{goetze91} 
         W. G\"otze in 
         {\it Liquids, Freezing and Glass Transition} 
         edited by J.P. Hansen , D. Levesque D, and J. Zinn-Justin  
         (Amsterdan: North-Holland) p~287, 1991.

\bibitem{goetzepisa}
           W. G\"otze, {\it J. Phys.: Condens. Matter}. {\bf 11}, A1 (1999).

\bibitem{hopping} S. P. Das and G. F. Mazenko, {\it Phys. Rev. A} {\bf
	34}, 2265 (1986); W. G\"otze and L. Sj\"ogren, {\it
	Z. Phys. B} {\bf 65}, 415 (1987).

\bibitem{pel} L. Angelani, R. Di Leonardo, G. Ruocco, 
	A. Scala, and F. Sciortino, {\it Phys. Rev. Lett.} {\bf 85}, 
	5356 (2000);T. S. Grigera, A. Cavagna, I. Giardina and G. Parisi,
		{\it Phys. Rev. Lett.} {\bf 88}, 055502 (2002).

\bibitem{aging} 
	W. Kob and J.-L. Barrat, {\it Phys. Rev. Lett.} {\bf 78}, 4581 (1997);
	L. F. Cugliandolo, condmat/0210312.

\bibitem{sciotar} F. Sciortino and P. Tartaglia, {\it J. Phys.:
Condens. Matter} {\bf 13}, 9127 (2001).



\bibitem{sperl} M. Sperl, {\it Phys. Rev. E} {\bf 68}, 031405 (2003).

\bibitem{konstanz} We have shown that Percus-Yevick approximation is
very accurate in the reentrant liquid region in E. Zaccarelli,
G. Foffi, K. A. Dawson, S. Buldyrev, F. Sciortino, P. Tartaglia, {\it
J. Phys.: Condens. Matter} {\bf 15}, 223 (2003).

\bibitem{zacca03} E. Zaccarelli, G. Foffi, F. Sciortino and
P. Tartaglia,  {\it Phys. Rev. Lett.} {\bf 91}, 108301 (2003).


\bibitem{notagamma} The values of $\gamma$ reported in
\protect\cite{zaccarelli02} are consistely higher than those of
Fig. \protect\ref{fig:fits}, due to the less numerical points
available there and consequently higher uncertainty in the fits.

\bibitem{foffialpha} G. Foffi, W. G\"otze, F. Sciortino, P. Tartaglia
and T. Voigtmann, {\it Phys. Rev. E} {\bf 69}, 011505 (2004).

\bibitem{saika04} I. Saika-Voivod, E. Zaccarelli, F. Sciortino and
P. Tartaglia, submitted to Phys. Rev. E (2004).




\end{thebibliography}
\end{document}